
\magnification=1200
\hsize=28pc
\vsize=43pc

\font\fg=eufm10 

\font\fontc=cmr10 scaled\magstep2

\font\fontg=cmr10

\newcount\notenumber

\def\cnote#1{\global\advance\notenumber by 1 \eqno(#1.\the\notenumber)}
\def\note{\global\advance\notenumber by 1 \eqno(\the\notenumber)}

\def\o{\omega}
\def\O{\Omega}
\def\d{\delta}
\def\pd{\partial}
\def\g{\gamma}
\def\si{\quad}
\def\sii{\qquad}

\footline={\hfill-- \folio\ --\hfill}

\rightline{TMUP-HEL-9306}
\rightline{June, 1993}
\vglue 1.0cm
\centerline{\fontc Large-N Collective Field Theory Applied to}
\centerline{\fontc Anyons in Magnetic Fields}
\vglue 3.0cm
\centerline{\fontg Hideaki Hiro-Oka\footnote{\dag}{e-mail:
hiro-oka@phys.metro-u.ac.jp} and Hisakazu Minakata\footnote{\ddag}{e-mail:
minakata@phys.metro-u.ac.jp}}
\vglue 0.5cm
\centerline{\it Department of Physics}
\centerline{\it Tokyo Metropolitan University }
\centerline{\it 1-1 Minami-Osawa Hachioji, Tokyo 192-03, Japan}
\vglue 3.0cm
\centerline{\fg Abstract}
\vglue 0.5cm
We present a large-$N$ collective field formalism for anyons in external
magnetic fields interacting with arbitrary two-body potential. We discuss how
the Landau level is reproduced in our framework. We apply it to the soluble
model for anyons proposed by Girvin et al., and obtain the dispersion relation
of collective modes for arbitrary statistical parameters.

\vfill\eject


\vglue 0.5cm


The large-$N$ collective field theory [1] has been successfully applied to a
wide variety of problems, the plasma oscillations, matrix models, and the
fractional quantum Hall system [2,3]. In this paper we attempt to extend its
applicability to the system of fractional statistics particles, anyons [4], in
magnetic fields. Since this system is closely related to the quantum Hall
system [5] including hierarchical excited states, it may be meaningful to
develop its treatment by the collective field method.

We first extend the collective field formalism so as to include gauge fields.
We believe that our method is one of the first explicit realizations of the
collective field theory including gauge fields. At the end of this paper we
shall make some comments on the similar approach taken in the literature [6].

We shall start by introducing our method.
For definiteness we shall work with the following simple $N$-body Hamiltonian
with arbitrary gauge fields $A$ in $2+1$ dimensions
$$
H={1\over 2}\sum^N_{a=1}\biggl(-i\pd_a+A(x_a)\biggr)^2, \note
$$
with the gauge condition $\pd A=0$. The mass $m$ and the coupling constant $e$
are set to unities throughout this paper to simplfy our equations.
The convensional way of treating this system is to solve the many-body Schr{\"
o}dinger equation
$$
H\psi(x_1,\cdots,x_N)=E\psi(x_1,\cdots,x_N).\note
$$

We treat this system by using the large-$N$ collective field formalism.
Following by now familiar method we shall rewrite above Hamiltonian into the
effective Hamiltonian in terms of collective variables.
We introduce a collective variable $Q^a$ by defining the point canonical
transformation
$$
x^a\longrightarrow Q^a=f^a(x),\note
$$
and its inverse transformation
$$
x^a=F^a(Q).\note
$$
By the chain rule of differentiation the Hamiltonian (1) is written as
$$
H={1\over 2}\sum^N_{a=1}\biggl(i\o^a
P_a+\sum^N_{b=1}\O^{ab}P_aP_b-\g^aP_a\tilde A-2\tilde A\g^aP_a+\tilde
A^2\biggr),\note
$$
where
$$
\o^a=-\sum^N_{b=1}{{\pd^2f^a}\over{\pd {x^b}^2}},\si P_a={\pd \over{i\pd
Q^a}},\si
\O^{ab}=\sum^N_{c=1}{{\pd f^a}\over {\pd x^c}} {{\pd f^b}\over {\pd x^c}},\note
$$
$$
\g^a=\sum^N_{b=1}{{\pd Q^a}\over {\pd x^b}},\si \tilde A(Q)=A(F(Q)).\note
$$
This expression (5) itself is not hermitian under the following na\"\i ve
hermitian conjugation
$$
{P_a}^\dagger=P_a,\sii {Q^a}^\dagger=Q^a.\note
$$
As is well understood this does not mean any trouble. The Hamiltonian which is
hermitian in the functional space of $\sqrt{J}\psi$ is the effective
Hamiltonian
$$
H_{eff}=J^{1/2}HJ^{-1/2},\note
$$
where $J$ is the Jacobian of the transformation (3). We define a new variable
$C$ as
$$
\eqalign{J^{1/2}P_aJ^{-1/2}&=P_a+iC_a,\cr
C_a&={1\over 2}{\pd\over {\pd Q}}\ln J,\cr}\note
$$
in order to bypass the difficulty of evaluating Jacobian. Substituting (10)
into the expression (9) and imposing the hermiticity condition
$H_{eff}^\dagger-H_{eff}=0$, we obtain the following equation for $C$
$$
\eqalign{\sum_{a=1}^N&{\pd\over {\pd Q^a}}\biggl\{\o^a +
\sum_{b=1}^N\Bigl(2\Omega^{ab} C_b +{\pd\Omega^{ab}\over {\pd
Q^b}}\Bigr)\biggr\}\cr
&+i\sum_{a=1}^N\Biggl[2\biggl\{\o^a + \sum_{b=1}^N\Bigl(2\Omega^{ab} C_b
+{\pd\Omega^{ab}\over {\pd Q^b}}\Bigr)\biggr\}P_a
+2\tilde A\Bigl(2\gamma^aC_a-{{\pd\gamma^a}\over{\pd
Q^a}}\Bigr)\Biggr]=0.\cr}\note
$$
The real and imaginary parts of left hand side in (11) should vanish
independently. Then one might worry that $C_a$ is overdetermined. But it is not
the case. One can show that the solution of the equation
$$
\o^a+\sum^N_{b=1}\biggl(2\O^{ab}C_b+{\pd\O^{ab}\over {\pd Q^b}}\biggr)=0,\note
$$
simultaneously solves the equation of vanishing imaginary part in the case of
the transformation
$$
\phi (x)={1\over N}\sum^N_{i=1}\d(x-x_i),\si \phi_k=\int dx
e^{-ikx}\phi(x).\note
$$
We will describe a short proof of this statement in Appendix. We note that this
feature should be generically true for any choice of the collective variable
besides (13). If it were not for the case we would have to make an ingenious
choice of the gauge fields so that the effective Hamiltonian becomes hermitian.

We utilize the density variable $\phi$ in (13) as an appropriate collective
variable to discuss the low energy collective excitation of the system. Since
we only need to solve (12) and it is the same equation as that without gauge
fields we can just go straight ahead as in Ref.2.

Solving (12) for $C$ and putting it into (9) we can obtain a complete effective
Hamiltonian up to a total divergence term. The Hamiltonian in terms of the
collective variable $\phi$ and its conjugate momentum $\pi$ is finally written
as
$$
H_{eff}={1\over 2}\int dx \biggl({{\pd\pi}\over
N}+A\biggr)N\phi\biggl({{\pd\pi}\over N}+A\biggr)+{N\over 8}\int
dx{{(\pd\phi)^2}\over \phi},\note
$$
The conjugate variable $\pi$ is defined as $\pi_k=\displaystyle{{1\over
i}{\pd\over {\pd \phi_{-k}}}}$ in momentum space and satisfies the commutation
relation
$$
\bigl[\ \phi(x),\pi(x')\ \bigr]=i\delta(x-x'),\note
$$
in the large $N$ limit.
In this way we have obtained the effective Hamiltonian with gauge field
expressed by the collective variables.


In the following we analyze the collective motion of anyons by means of the
collective field formalism just developed. Instead of working with ideal anyon
Hamiltonian (1) with Schr{\" o}dinger's wave function obeying fractional
statistics we make a singular gauge transformation to have an interacting boson
representation of the ideal anyon gas. The Hamiltonian is given as
$$
H=\sum^N_{a=1}{1\over 2}\biggl(-i\pd_a+a(x_a)\biggr)^2,\note
$$
where $a$ is the so-called fictitious gauge field written as
$$
a_i(x_a)={\theta\over\pi}\sum^N_{b\ne
a}{{\epsilon_{ij}(x_a-x_b)^j}\over{|x_a-x_b|^2}},\sii \theta\in [0,\pi],\note
$$
where the spatial index $i$ is explicitly indicated.
The parameter $\theta$ interpolates the two special limits, $\theta\rightarrow
0$ (boson) and $\theta\rightarrow\pi$ (fermion).

We consider, for later convenience, more generic case of anyons in real
(opposed to fictitious) magnetic field interacting with each other by the
potential $U$. The Hamiltonian is given by
$$
H=\sum^N_{a=1}{1\over 2}\biggl(-i\pd_a +a(x_a)+A(x_a)\biggr)^2+\sum_{a<b}^N
U(x_a-x_b),\note
$$
where $A$ is the real gauge field yielding the external constant magnetic field
$B$
by $\nabla\times A=B$,
and $U$ is the two body interaction whose explicit form will be specified
later.

Following the same procedure as before it is easy to derive the effective
Hamiltonian corresponding to that of (18) to the following,
$$
\eqalign{H={1\over 2}&\int dx \biggl({{\pd\pi}\over
N}+a+A\biggr)N\phi\biggl({{\pd\pi}\over N}+a+A\biggr)+{N\over 8}\int
dx{{(\pd\phi)^2}\over \phi}\cr
+&{N^2\over 2}\int dxdy\biggl(\phi-{1\over V}\biggr) U(x-y)\biggl(\phi-{1\over
V}\biggr)+\lambda\biggl(\int dx\phi-1\biggr).\cr}\note
$$
where
$$
a_i(x)={\theta\over \pi}N\int
dy{{\epsilon_{ij}(x-y)^j}\over{|x-y|^2}}\phi(y).\note
$$
is the corresponding expression of (17) in terms of the density variable. $V$
is the volume of the system.
The last term in (19) is due to the constraint $\int dx\phi=1$ and $\lambda$ is
the Lagrange multiplier (chemical potential).

The ground state of this system can be given by solving the following equations
$$
\nabla\biggl\{N\phi\biggl({{\nabla\pi}\over N}+a+A\biggr)\biggr\}=0,\note
$$
$$
\eqalign{
-{N\over 8}&\biggl({{\nabla\phi}\over\phi}\biggl)^2-{N\over 4}\nabla
\biggl({{\nabla\phi}\over\phi}\biggr)+{\theta\over\pi}N^2\int dy \phi\nabla \ln
|x-y|\times\biggl({{\nabla\pi}\over N}+a+A\biggr)\cr
&+{N\over 2}\biggl({{\nabla\pi}\over N}+a+A\biggr)^2+{N^2\over 2}\int dyU(x-y)
\biggl(\phi(y)-{1\over V}\biggr)-\lambda=0\cr}\note
$$
Assuming that the ground state has no vortex and that it is invariant under the
translation, the solution can be given as
$$
\phi_0={1\over V}.\note
$$
Moreover, we notice that
$$
{{\nabla\pi_0}\over N}+a_0+A=0\note
$$
satisfies these equations (21) and (22) under the solution (23). Here $a_0$ is
the mean field defined by
$$
a_i(x)=a_{0i}(x)+{\theta\over\pi}N\int
dy{{\epsilon_{ij}(x-y)^j}\over{|x-y|^2}}\biggl(\phi(y)-{1\over V}\biggr).\note
$$
As a solution of the equation (24) we take
$$
\pi_0=0,\note
$$
and
$$
a_0+A=0.\note
$$

The latter equation means that the fictitious megnetic field organizes itself
so as to cancell the applied real magnetic field. If the magnetic fields would
not cancell with each other, we would have to have a singularity in $\pi$
variable. This is easily seen by (24) which states that
$\nabla\times(\nabla\pi_0)\ne 0$ unless $\nabla\times(a_0+A)=0$. The
singularity of $\pi$ variable implies a real vortex. Therefore, our solution to
(24) means the absense of extra real vortices.

The validity of taking the above solution to (24) can be explicitly checked.
The second equation (27) implies that the ground state value $\phi_0$ is
determined by the external magnetic field $B$. That is, the ground state
density $\rho_g$ is given as
$$
\rho_g={B\over{2\theta}}.\note
$$

For fermions ($\theta=\pi$) this gives $\rho_g=B/2\pi$, the well-known result
for the degeneracy of the lowest Landau level. Therefore, our solution (27)
allows a very simple interpretation of completely filled lowest Landau level in
this case. It is amusing to observe that the density (28) interporates between
the Landau level degeneracy for the fermion case ($\theta=\pi)$ and the Bose
condensate for the bosonic case $(\theta\rightarrow 0)$.

Before moving to the analysis of collective excitations let us make a short
comment on ideal anyon gas without external magnetic field. In the absence of
$A$ in (24) we have to deal with the singular configuration of the $\pi$ field,
the Chern-Simons vortex. In this case the $\phi$ field must vanish at the
location of vortices and we are not allowed to have a constant solution (23).
Therefore the treatment of the ideal anyon gas is far from straightforward in
the large-$N$ collective field theory. We hope to return to it in the future.

We analyze collective excitations as small oscillations around the ground state
given by (26) and (27). We expand the $\phi$ and the $\pi$ variables around
their ground state values,
$$
\phi=\phi_0+{1\over\sqrt{N}}\eta,\sii\pi=\pi_0+\sqrt{N}\xi.\note
$$
The prefactors of the fluctuation $1/\sqrt{N}$ and $\sqrt N$ are determined in
consistent with the large-$N$ expansion and the canonical commutation relation.
As a mode decomposition we expand $\eta$ and $\xi$ by using the plane wave,
{\it i.e.}
$$
\eqalign{
\eta=&{1\over V}\sum_{k\ne 0}\sqrt{k^2}e^{ikx}q_k,\cr
\xi=&\sum_{k\ne 0}{1\over\sqrt{k^2}}e^{-ikx}p_k,\cr}\note
$$
Inserting equations (23), (26), and (30) into the Hamiltonian (19) and picking
up the $O(N^0)$ terms, we obtain the simple harmonic oscillator type
Hamiltonian
$$
H_{N^0}=\sum_{k\ne 0}{1\over 2}\biggl(|p_k|^2+\o_k^2|q_k|^2\biggr),\note
$$
where
$$
\o_k^2={k^4\over 4}+4\theta^2\rho^2+\rho k^2U_k,\note
$$
and $U_k$ is the Fourier components of $U(x-y)$.
This is the central result of our paper.

Let us make a few remarks:
\vglue 0.2cm
\item{(i)} The first term, which dominates in short wavelength limit, $k\gg 1$,
is the familiar kinetic term of non-relativistic particles. In this limit the
information of the fractional statistics is lost as one can expect.
\item{(ii)} The second term, which is of order $\hbar^0$, dominates in the
classical region (or in the long wavelength limit). This is the gap that can be
interpreted as the Landau level interval. To see this we note that $\rho$ is
given by (28) for the ground state. Then $\o=B$ which is nothing but the
cyclotron frequency under our choice of the units $m=1$ and $e/c=1$. It is also
natural that the cyclotron frequency is independent of the $\theta$ parameter.
Thus our formalism passes the test [7] using Kohn's theorem [8].
\vglue 0.2cm
Let us discuss a soluble model of fractional statistics [9,10] as a special
example of our generic Hamiltonian (18). In this model the potential $U$ is
given by
$$
U(x-y)=2\theta\delta^{(2)}(x-y)\note
$$
In this case the Fourier component is given as $U_k=2\theta$ and the energy
level given by (32) simplifies:
$$
E_k=\hbar\o_k={k^2\over 2}+2\theta\rho.\note
$$
This result at $\theta=\pi/2$ (semion case) is in agreement with the one
derived by Girvin et al. [9] using the plasma analogy. Our result generalizes
it to arbitrary $\theta$ cases.


Next we shall consider the Coulomb interaction
$$
U(x-y)={1\over{\epsilon|x-y|}},\note
$$
where $\epsilon$ is the background dielectric constant. The Fourier component
$U_k$ is given by
$$
U_k={2\pi\over {\epsilon k}}.\note
$$
In this case the frequency (32) becomes, in long wavelength limit ($k\ll 1$),
as
$$
\o_k=B+{\pi\over{2\theta\epsilon}}k.\note
$$
This agrees with the result of RPA calculation by Kallin and Halperin [11] at
$\theta=\pi$ (fermion), and reproduces that of Zhang [7] for general $\theta$
cases derived by using the density-density correlation function.


A closely related but different collective field formalism of anyon system has
been proposed by Andri{\'c} and Bardek [6]. While they employ the same
technology as ours two formalisms differ because the starting Hamiltonians
differ. Their Hamiltonian is supposed to act to the symmetric wave function
constructed by removing an antisymmetrized factor. As a consequence it is
non-hermitian. If we repeat the similar fluctuation analysis using their
collective field Hamiltonian, eq. (16) in Ref.6, we obtain as $\o_k^2$
$$
\o_k^2=\o_k^2(\hbox{ours})+{\theta\over {4\pi}}\biggl({\theta\over
{4\pi}}-1\biggr)k^4-{\theta\over {\pi}}\biggl({\theta\over
{2\pi}}-1\biggr)2\pi\rho k^2.\note
$$
Therefore, two formalisms differ also in physical results.
\vglue 0.5cm

To summarize: We have constructed a large-$N$ collective field theory for
anyons in external magnetic fields interacting with an arbitrary two-body
potential. We illuminated how the Landau level is reproduced in our framework
and clarified the basic picture behind this, namely, the cancellation between
the real and the fictitious magnetic fields. We applied our collective field
formalism to uncover the low energy collective excitations around the ground
state. In particular we obtained the dispersion relation for the collective
mode in the soluble model of anyons for generic value of the statistical
parameter $\theta$.

\vglue 0.5cm
{\bf Acknowledgments}

The authors thank Bunji Sakita for numerous discussions on the collective field
theory and for the hospitality accorded to them during a visit to City College
of New York.

\vfill\eject
{\bf Appendix}

In this Appendix we show that the left hand side of (11) vanishes under the
choice of the collective variable (13) when (12) is satisfied. Namely, we shall
show that
$$
2\tilde A\Bigl(2\gamma^aC_a-{{\pd\gamma^a}\over{\pd Q^a}}\Bigr)\eqno(A1)
$$
vanishes under the above conditions. For convenience, we consider the Fourier
component hereafter. From (12) we have
$$
C_k={1\over 2}\sum_{k'}\biggl(\Omega^{kk'}\biggr)^{-1}\o^{k'}.\eqno(A2)
$$
We note that under the choice (13), namely $Q^a=\phi_k$,
$$
\gamma_k=-ik\phi_k\eqno(A3)
$$
and
$$
\Omega^{kk'}=-{{kk'}\over N}\phi_{k+k'}.\eqno(A4)
$$
Using (A3) one can show that the second term of (A1) is propotional to
$\sum_kk$ and vanishes due to the symmetry $k\leftrightarrow -k$. Whereas the
first term of (A1) can be written, thanks to (A3) and (A4), as
$$
\eqalign{
\sum_k2\gamma^kC_k=&
\sum_{kk'}ik'\phi_k\phi_{k'}\phi_{k+k'}^{-1}\cr
=&\sum_{kk'}\int dx_1dx_2dx_3\
ik'e^{-ikx_1}\phi(x_1)e^{-ik'x_2}\phi(x_2)e^{i(k+k')x_3}\phi^{-1}(x_3)\cr
=&\sum_{kk'}\int dx_1dx_2dx_3\
\pd\phi(x_1)\phi(x_2)\phi^{-1}(x_3)e^{-ik(x_1-x_3)}e^{-ik'(x_2-x_3)}\cr
=&\int dx\ \pd\phi(x)\cr
=&0,}
$$
unless some topologically nontrivial configurations arise.
Thus we have shown that (A1) vanishes when (12) is satisfied.

\vskip 1cm
{\bf References}

\item{1.} A. Jevicki and B. Sakita, Nucl. Phys. B165 (1980) 511, B185 (1981)
89.
\item{2.} B. Sakita, Quantum Theory of Many-Variable Systems and Fields (World
Scientific, Singapore, 1985).
\item{3.} B. Sakita, D.-N. Sheng, and Z.-B. Su, Phys. Rev. B44 (1991) 11510.
\item{4.} For a summary and the original references on anyons, see e.g., F.
Wilczek, Fractional Statistics and Anyon Superconductivity (World Scientific,
Singapore, 1990).
\item{5.} R. E. Prange and S. M. Girvin (ed.), The Quantum Hall Effect, 2nd
edition (Springer-Verlag, Berlin, 1989).
\item{6.} I. Andri{\'c} and V. Bardek, Mod. Phys. Lett. A7 (1992) 3276.
\item{7.} S. C. Zhang, Int. J. Mod. Phys. B6 (1992) 25.
\item{8.} W. Kohn, Phys. Rev. 123 (1961) 1242.
\item{9.} S. M. Girvin, A. H. MacDonald, M. P. A. Fisher, S.-J. Rey, and J. P.
Sethna, Phys. Rev. Lett. 65 (1990) 1671.
\item{10.} M. Greiter and F. Wilczek, Nucl. Phys. B370 (1992) 577.
\item{11.} C. Kallin and B. Halperin, Phys. Rev. B30 (1984) 5655.

\end